\documentclass[a4paper,11pt]{article}
\usepackage{amsmath,amsthm,latexsym,amssymb,amsfonts,epsfig}
\usepackage{fontenc,dsfont,layout}
\usepackage{hyperref}
\usepackage{psfrag}
\usepackage[cp1250]{inputenc}
\usepackage{graphicx}
\numberwithin{equation}{section}

\DeclareMathAlphabet{\mathpzc}{OT1}{pzc}{m}{it}

\begin{document}

\title{Saddle point inflation from $f(R)$ theory}
\author{Micha{\l} Artymowski$^1$\thanks{Michal.Artymowski@uj.edu.pl} $\;\,$ ,  $\;$
           Zygmunt Lalak$^2$\thanks{Zygmunt.Lalak@fuw.edu.pl} $\;\,$ and  $\;$ Marek Lewicki$^2$\thanks{Marek.Lewicki@fuw.edu.pl}}
\date{\it 1. Institute of Physics, Jagiellonian University\\
{\L}ojasiewicza 11, 30-348 Krak{\'o}w, Poland\\
2. Institute of Theoretical Physics, Faculty of Physics, University of Warsaw\\
 ul. Pasteura 5, 02-093 Warsaw, Poland} 
\maketitle

\abstract{{We analyse several saddle point inflationary scenarios based on power-law $f(R)$ models. We investigate inflation resulting from $f(R) = R + \alpha_n M^{2(1-n)}R^n + \alpha_{n+1}M^{-2n}R^{n+1}$ and $f(R) = \sum_n^l \alpha_n M^{2(1-n)} R^n$ as well as $l\to\infty$ limit of the latter. In all cases we have found relation between $\alpha_n$ coefficients and checked consistency with the PLANCK data as well as constraints coming from the stability of the models in question.
Each of the models provides solutions which are both stable and consistent with PLANCK data, however only in parts of the parameter space where inflation starts on the plateau of the potential, some distance from the saddle. And thus all the correct solutions bear some resemblance to the Starobinsky model.
}}

\maketitle

\section{Introduction}

Cosmic inflation \cite{Lyth:1998xn,Liddle:2000dt,Mazumdar:2010sa} is a theory of the early universe which predicts cosmic acceleration and generation of seeds of the large scale structure of the present universe. It solves problems of classical cosmology and it is consistent with current experimental data \cite{Ade:2013uln}. The first theory of inflation was the Starobinsky model \cite{Starobinsky:1980te}, which is an $f(R)$ theory \cite{DeFelice:2010aj} with $R + R^2/6M^2$ Lagrangian density. In such a model the acceleration of space-time is generated by the gravitational interaction itself, without a need to introduce any new particles or fields. The embedding of Starobinsky inflation in no-scale SUGRA has been discussed in Ref. \cite{Ellis:2015xna}. Recently the whole class of generalisations of the Starobinsky inflation have been discussed in the literature \cite{Codello:2014sua,Ben-Dayan:2014isa,Artymowski:2014gea,Artymowski:2014nva,Sebastiani:2013eqa,Motohashi:2014tra,vandeBruck:2015xpa}, also in the context of the higher order terms in Starobinsky Jordan frame potential \cite{Broy:2014sia,Kamada:2014gma,Artymowski:2015mva}.
\\*

The typical scale of inflation is set around the GUT scale, which is of the order of $(10^{16}GeV)^4$. Such a high scale of inflation seems to be a disadvantage of inflationary models. First of all inflationary physics is very far away from scales which can be measured in accelerators and other high-energy experiments. The other issue is, that high scale of inflation enables the production of super-heavy particles during the reheating \cite{Mukhanov:2005sc}. Those particles could be in principle heavier than the inflaton itself, so particles like magnetic monopoles, which abundant existence is  inconsistent with observations, could be produced after inflation. Another argument, which supports low-scale inflation is the Lyth bound \cite{Lyth:1996im}, which is the relation between variation of the inflaton during inflation in Planck units (denoted as $\Delta\phi$) and tensor-to-scalar ratio $r$, namely
\begin{equation}
\Delta\phi \simeq \int_0^N\sqrt{\frac{r}{8}}dN \, ,
\end{equation}
which for nearly scale-invariant power spectrum gives $\Delta\phi<M_p$ for $r<0.002$. Small $\Delta\phi$ seems to be preferable from the point of view of the naturalness principle, since $M_p$ is the cut-off scale of the theory. The value of $r$ determines the scale of inflation, since $V/r$ (where $V$ is the potential of the inflaton) at the scale of inflation is set by the normalisation of CMB anisotropies. Therefore in order to obtain small $r$ one needs a low-scale inflation, which may be provided by a potential with a saddle point.
\\*

A separate issue related with $f(R)$ inflation is related with loop corrections to the f(R) function. In order to obtain quasi de Sitter evolution of space-time one needs a range of energies for which the $R^2 M^{-2}$ term dominates the Lagrangian density. This would require all higher order corrections (such as $R^3$, $R^4$ etc.) to be suppressed by a mass scale much bigger than $M$. One naturally expects all higher order correction to GR to appear at the same energy scale if one wants to avoid the fine-tuning of coefficients of all higher order terms. From this perspective it would be better to generate inflation in $f(R)$ theory without the Starobinsky plateau, which in principle could be obtained in the saddle point inflation.
\\*

In {what follows} 
 we use the convention $8\pi G = M_{p}^{-2} = 1$, where $M_{p}\sim 2\times 10^{18}GeV$ is the reduced Planck mass.
\\*

The {outline of the paper} 
 is as follows. In Sec. \ref{sec:intro} we give short introduction to $f(R)$ and its description as a Brans-Dicke theory. In Sec. \ref{sec:FR} we discuss three saddle point $f(R)$ scenarios, namely: i) two higher order terms $R^n$ and $R^{n+1}$, ii) at least 4 higher order terms with powers bigger than 2, iii) infinite number of higher order terms with finite sum at every energy scale. Finally we summarise in Sec. \ref{sec:concl}


\section{Introduction to $f(R)$ theory, inflation and primordial inhomogeneities} \label{sec:intro}

The $f(R)$ theory is one of the simplest generalisations of general relativity (GR). It is based on Lagrangian density $S = \frac{1}{2}\int d^4 \sqrt{-g}f(R)$ and it can be expressed using the so-called auxiliary field $\varphi$ defined by $\varphi=F(R):=\frac{df}{dR}$. In such a case the Jordan frame (JF) action is equal to $S = \int \sqrt{-g} (\varphi R/2 - U(\varphi)$, where $U = (RF-f)/2$ is the JF potential. For $F=1$ one recovers GR, so the GR vacuum of the JF potential {is positioned at}
 $\varphi=1$. The same model can be expressed in the Einstein frame (EF), with the metric tensor defined by $\tilde{g}_{\mu\nu} = \varphi g_{\mu\nu}$. This is purely classical transformation of coordinates and results obtained in one frame are perfectly consistent with the ones from another frame \footnote{Differences between Einstein and Jordan frame in loop quantum cosmology are described in Ref. \cite{Artymowski:2013qua}.}. The EF action is equal to $S = \int \sqrt{-\tilde{g}}(\tilde{R}/2+(\partial_\mu \phi)^2/2 - V(\phi))$, where $\tilde{R}$, $\phi:=\sqrt{3/2}\,\log F$ and $V:=(RF-f)/(2F^2)$ are the EF Ricci scalar, field and potential respectively. The EF potential should have a minimum at the GR vacuum, which is {positioned at} $\phi = 0$. 
\\*

In the EF the {gravity}
 obtains its canonical form and {this is why}
  the EF is usually used for the analysis of inflation and generation of primordial inhomogeneities. The cosmic inflation {proceeds} 
   when both slow-roll parameters $\epsilon$ and $\eta$ are much smaller than unity{. These parameters are, as usual given by}
\begin{equation}
\epsilon = \frac{1}{2}\left(\frac{V_\phi}{V}\right)^2 \, , \qquad \eta = \frac{V_{\phi\phi}}{V} \, ,
\end{equation}
where $V_{\phi}$ and $V_{\phi\phi}$ are the first and the second derivative of the EF potential with respect to $\phi$. During inflation $\epsilon$ and $\eta$ can be interpreted as deviation from the de Sitter solution for FRW universe. During each Hubble time the EF scalar field produces inhomogeneous modes with an amplitude of the order of the Hubble parameter. From them and from the scalar metric perturbations one constructs gauge invariant curvature perturbations, which are directly related to cosmic microwave background anisotropies. Their power spectrum $\mathcal{P}_\mathcal{R}$, their spectral index $n_s$ and their tensor to scalar ratio are {as follows}
\begin{equation}
\mathcal{P}_\mathcal{R} \simeq \frac{V}{24\pi^2\epsilon}\, , \quad n_s \simeq 1- 6\epsilon + 2\eta\, , \quad r \simeq 16 \epsilon. 
\end{equation}
In the low scale inflation one obtains $\epsilon \ll |\eta|$, which for $\eta<0$ gives $1-n_s\simeq 2|\eta|$.


\section{Saddle point inflation in power-law $f(R)$ theory} \label{sec:FR}

\subsection{Saddle point with vanishing two derivatives} \label{sec:2terms}

\begin{figure}[h]
\centering
\includegraphics[height=5.7cm]{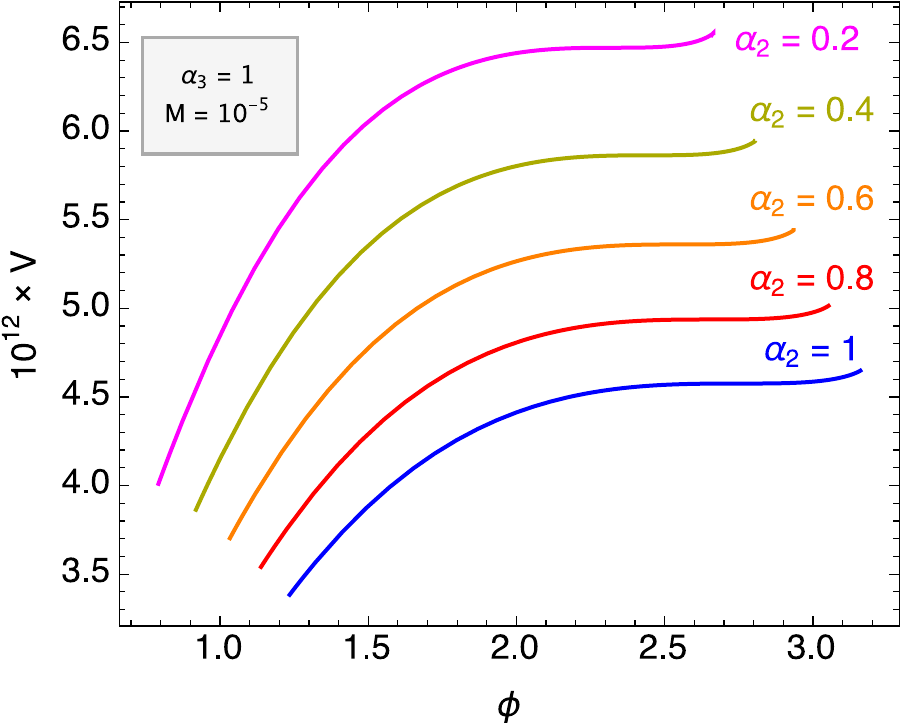}
\hspace{0.5cm}
\includegraphics[height=5.7cm]{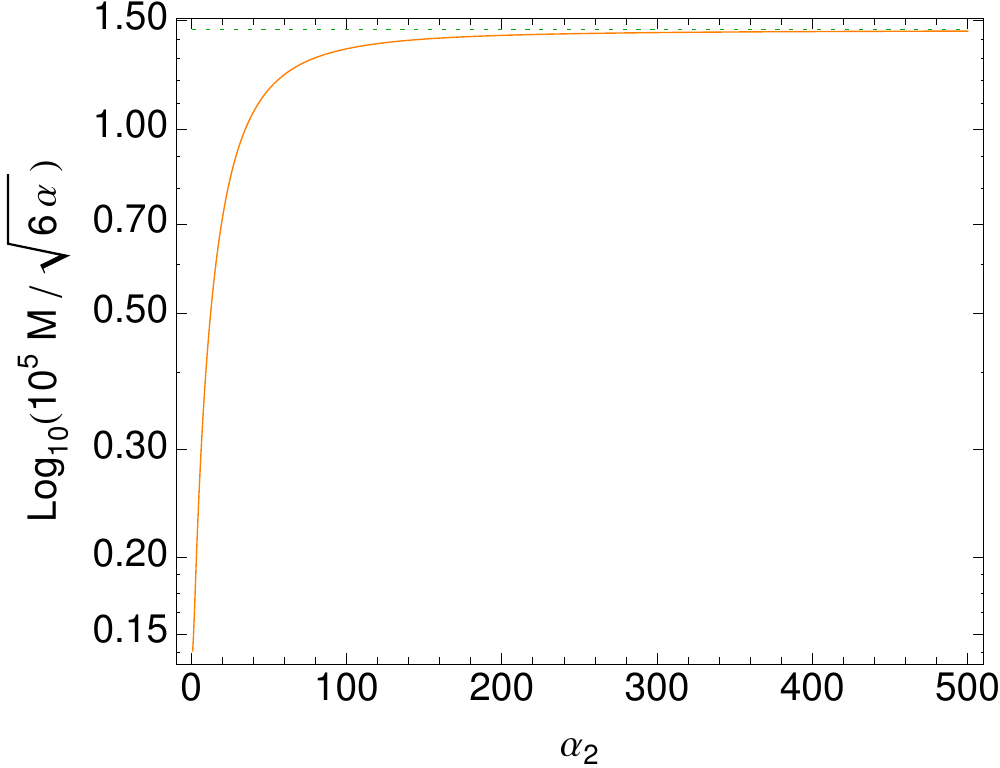}
\caption{\it Left panel: the Einstein frame potential for the (\ref{eq:Fsaddle}) model around the saddle point for $n=3$, $\alpha_3=1$ and different values of $\alpha_2$. The $\alpha_4$ coefficient is set from Eq. (\ref{eq:Rsalphas}) for $n=3$. The maximal allowed value of $\phi$ is very close to the saddle point. Right panel: the scale of new physics $M$ as a function of $\alpha_2$. Dotted green line represent Starobinsky limit.}
\label{fig:Vsaddle}
\end{figure}

As mentioned in the introduction, the loop corrections to the Starobinsky model {are} 
of the form $\sum_{n=2}^{\infty} \alpha_n R^n M^{2(1-n)}$, where $M$ is a mass scale, which suppresses deviations from GR. Therefore, in order to obtain sufficiently long Starobinsky plateau one needs a broad range of energy scales on which $R^2$ dominates over all higher order corrections. This requires a fine-tuning of infinite number of $\alpha_n$ coefficients. To avoid that we will consider an inflationary scenario in which different higher-order corrections can become relevant at the same energy scale, namely the saddle-point inflation from a power-law $f(R)$ theory. For general form of $f(R)$ one obtains a saddle point of the Einstein frame potential for $V_\phi = V_{\phi\phi} =0$, which corresponds to
\begin{equation}
RF=\frac	{1}{2}f \, , \qquad RF'=F \, , \label{eq:saddleFR}
\end{equation}
where $F = f'$ and prime denotes the derivative with respect to the Ricci scalar. Let us assume the following form of $f(R)$
\begin{equation}
f(R) = R + \alpha_2 \frac{R^2}{M^2} + \alpha_n \frac{R^{n}}{M^{2(n-1)}} + \alpha_{n+1}\frac{R^{n+1}}{M^{2n}} \, , \label{eq:Fsaddle}
\end{equation}
{where $n > 2$ is a given number. In such a case the saddle point appears for}
\begin{equation}
R = R_s = M^2\left(\frac{(n-2) \alpha_n }{n}\right)^{\frac{-1}{n-1}} \, , \qquad \alpha_{n+1} = - \left(\frac{(n-2) \alpha_n }{n}\right)^{\frac{n}{n-1}} \, .\label{eq:Rsalphas}
\end{equation}
Equations above are $\alpha_2$ independent, because any $R^2$ can satisfy Eq. (\ref{eq:saddleFR}). In order to keep $R_s$ and $\alpha_{n+1}$ real we need to assume that $\alpha_{n}>0$ and $\alpha_{n+1}<0$. Then for sufficiently big $R$ one finds $F<0$ and the gravity becomes repulsive. This instability becomes an issue for $R\simeq M^2\frac{\alpha_n  n}{n+1} \left(\alpha_n  (n-2)/n\right)^{\frac{-n}{n-1}}$, which is typically of the same order of magnitude as $R_s$. By redefining $M$ we can always set one of $\alpha_n$ to be any given constant. For negative $\alpha_n$ one can satisfy Eq. (\ref{eq:saddleFR}) for $n<2$. Nevertheless the saddle point would lie in the repulsive gravity regime, where $F<0$. Thus in the following analysis $n<2$ is excluded. Note that for non-zero value of $\alpha_2$ the value of $M$ grows with $\alpha_2$. This comes from the fact that for $\alpha_2\gg\alpha_n$ one obtains inflationary plateau followed by the saddle point, due to growing value of $R_s$ with respect to $\alpha_2$. The $\alpha_2$ dependence of $M$ is shown in Fig. \ref{fig:Vsaddle}. Big $\alpha_2$ term means that the last 60 e-folds of inflation happen on the Starobinsky plateau, so one does not obtain significant deviations from the $R^2$ model.
\\*

The Einstein frame potential around the saddle point (up to the maximal allowed value of $\phi$) for $f(R)=R + \alpha_2 R^2/M^2 +  R^3/M^4 + \alpha_4 R^4/M^6$ has been shown in Fig \ref{fig:Vsaddle}. We have rescaled $M$ to obtain $\alpha_3=1$. The $R^2$ term in not necessary to obtain a saddle point, but we include it to combine the inflation on the Starobinsky plateau with the saddle point inflation. From Eq. (\ref{eq:Rsalphas}) one finds the value of $\alpha_4$, normalisation of inhomogeneities gives $M$ as a function of $\alpha_2$.

\begin{figure}[h]
\centering
\includegraphics[height=5.5cm]{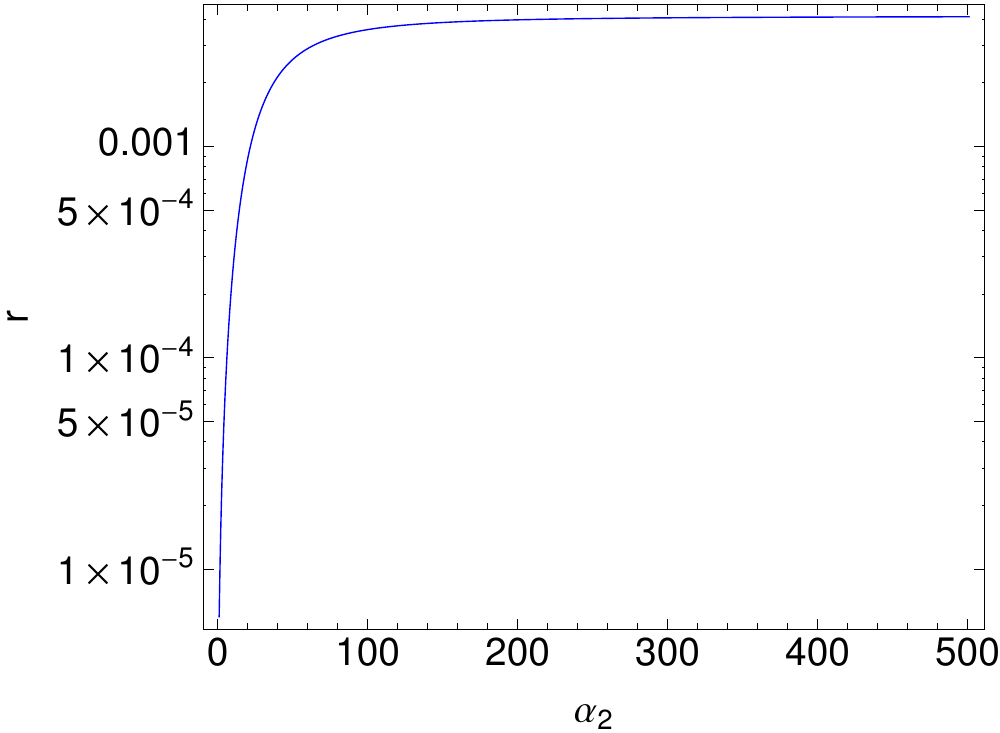}
\hspace{0.5cm}
\includegraphics[height=5.5cm]{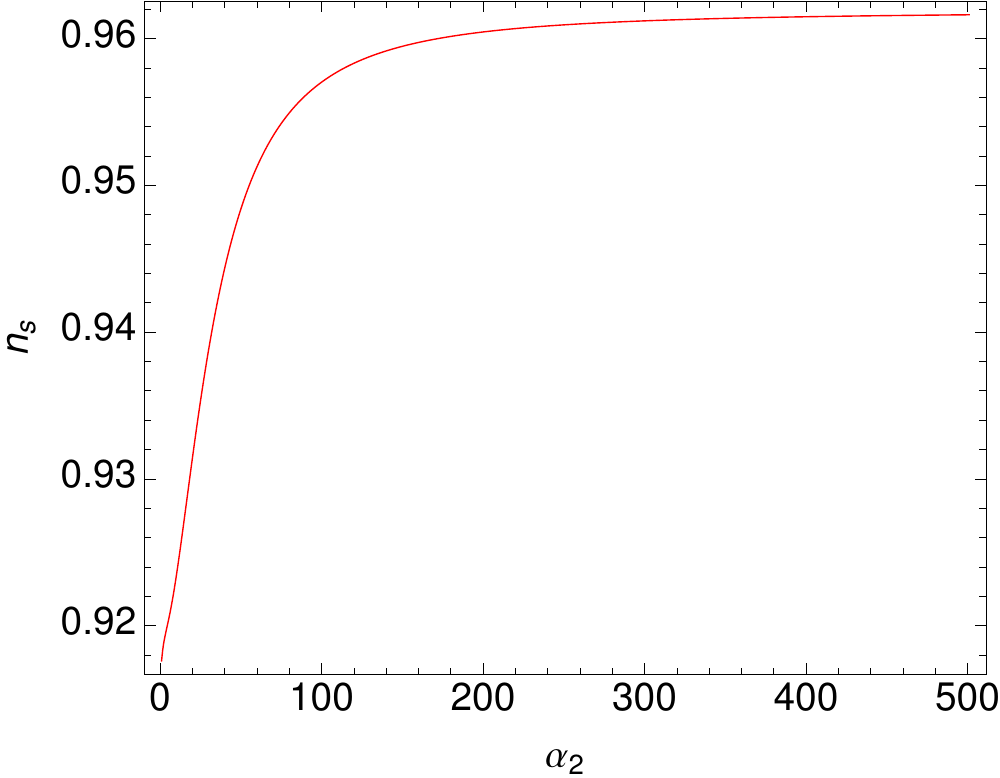}
\caption{\it {Tensor to scalar ratio $r$ and spectral index $n_s$ as a function of $\alpha_2$ for the (\ref{eq:Fsaddle}) model with $n=3$. One can fit the PLANCK data for $\alpha_2\gtrsim 100$, which means that the saddle point is preceded by the inflationary plateau.}}
\label{fig:rnsSaddle}
\end{figure}

\begin{figure}[h]
\centering
\includegraphics[height=5.5cm]{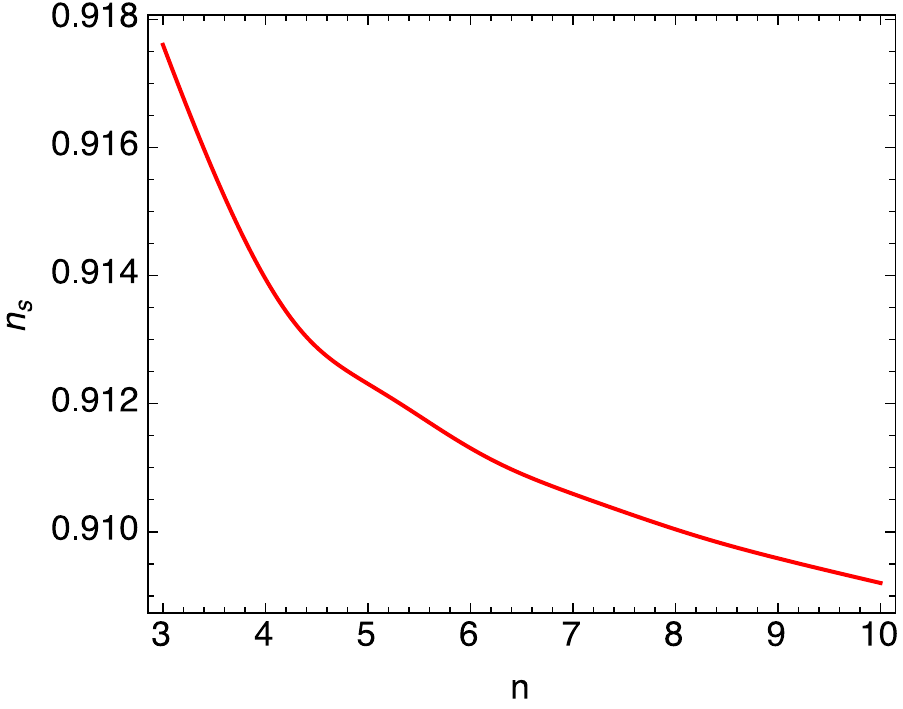}
\hspace{0.5cm}
\includegraphics[height=5.5cm]{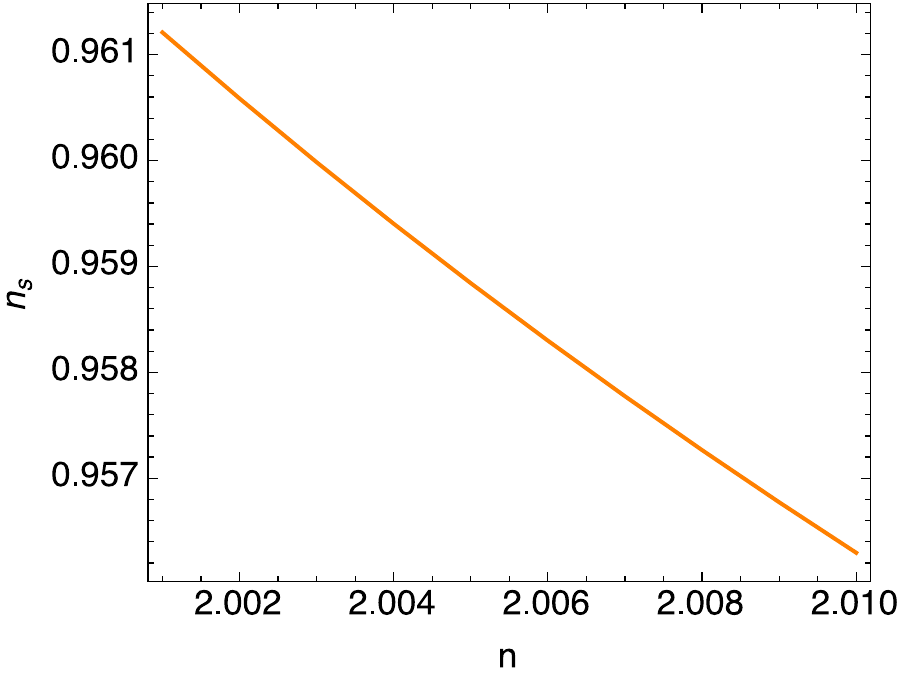}
\caption{\it {Both panels present $n_s(n)$ for the model from Eq. (\ref{eq:Fsaddle}) for $N_{\star}=50$ and $\alpha_2 = 0$. If $n$ is a natural number one cannot fit the Planck data due to too small $n_s$. In all of those cases a significant contribution of the $R^2$ term is needed in order to obtain $n_s \gtrsim 0.958$. On the other hand for $n-2\lesssim 10^{-2}$ one obtain $n_s\simeq 0.96$. The case of $n\gtrsim 2$ seems to be especially interesting since it allows to reconstruct Starobinsky results it the presence of higher order terms.}}
\label{fig:n_s(n)}
\end{figure}

\subsection{Saddle point with vanishing $k$ derivatives} \label{sec:kterms}

In general one can define the saddle point with first $k$ derivatives vanishing, which was analysed in Ref. \cite{Hamada:2015wea}. In {that} case $1-n_s \simeq \frac{2k}{N_{\star}(k-1)}$ when freeze-out of primordial inhomogeneities happens close to the saddle point. Thus, for sufficiently big $k$ one can fit the Planck data. In our case all $\frac{d^k V}{d\phi^k}=0$ at the saddle point are equivalent to $\frac{d^k f}{dR^k} = 0$ for $k>2$. The $f(R)$ model from Eq. (\ref{eq:Fsaddle}) cannot satisfy these equations, so in order to obtain a saddle point with vanishing higher order derivatives one needs to introduce more terms to $f(R)$ function. Thus let us {now consider} 
\begin{equation}
f(R) = R +\alpha_2 \frac{R^2}{M^2} + \sum _{n=3}^{l}\alpha_n\frac{R^n}{M^{2(n-1)}} \, ,\label{eq:sumF}
\end{equation}
{where $l>4$ is an even natural number. Again, without any loss of generality one can choose $\alpha_3$ to be any positive constant, so for simplicity we set $\alpha_3=1$. Then one can satisfy Eq. (\ref{eq:saddleFR}) and $f^{(n)}=0$ (for $n=\{3,4,\ldots,l-2\}$ and any value of $\alpha_2$) and the saddle point appears at}
\begin{equation}
R = {\bf R}_s = \sqrt{p}\, M^2 \, ,\qquad \text{where}\qquad p= \sqrt{(l-1) \left(\frac{l}{2}-1\right)} \, .\label{eq:bfRs}
\end{equation}
{The $\alpha_n$ coefficients satisfy}
\begin{equation}
\alpha_n = (-1)^{n-1}\frac{2(l-3)!}{(l-n)!(n-1)!}p^{\frac{3-n}{2}}   \quad \text{for} \quad n = \{3,\ldots,l\} \, . \label{eq:alphan}
\end{equation}
Note that Eq. (\ref{eq:bfRs}) and (\ref{eq:alphan}) are completely independent of $\alpha_2$. Since $\alpha_l < 0$ one obtains $F<0$ for sufficiently big $R$. Alike the model from Eq. (\ref{eq:Fsaddle}) the biggest allowed value of $R$ is slightly bigger than ${\bf R}_s$. Using Eq. (\ref{eq:sumF}) and (\ref{eq:alphan}) one obtains
\begin{equation}
f(R) = R+\frac{\alpha_2}{M^2} R^2+R\frac{\left(l M^2 \sqrt{p} R+2 M^4 p^2 \left(\left(1-\frac{R}{M^2 \sqrt{p}}\right)^l-1\right)-(l-1) R^2\right)}{M^4 p-M^2 \sqrt{p} R}\, . \label{eq:fSum}
\end{equation}

\begin{figure}[h]
\centering
\includegraphics[height=5.3cm]{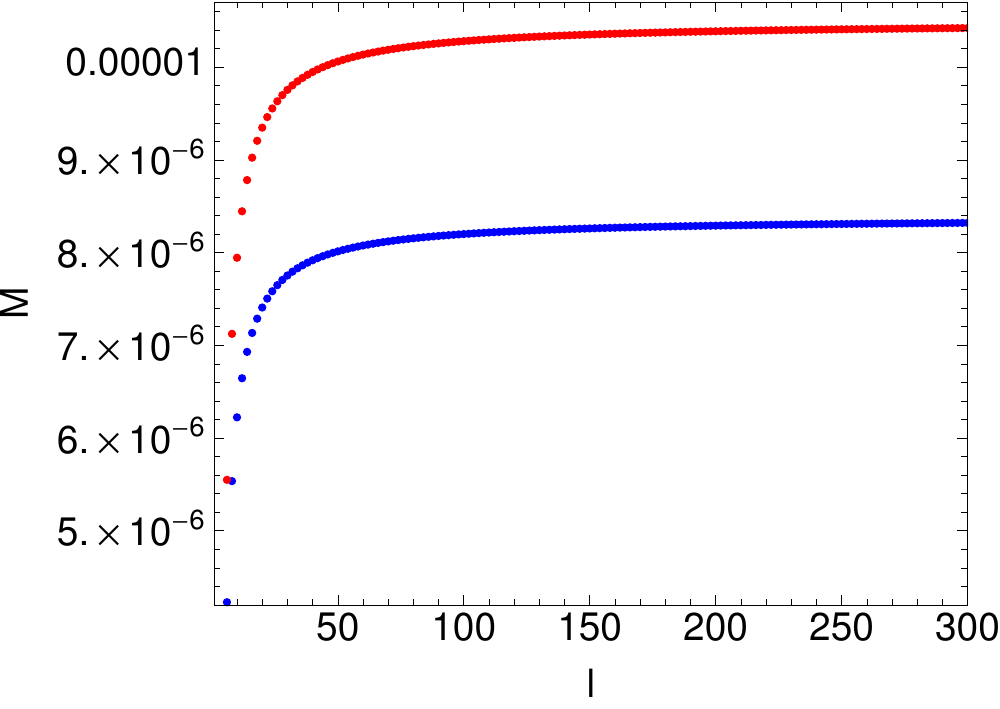}
\hspace{0.5cm}
\includegraphics[height=5.3cm]{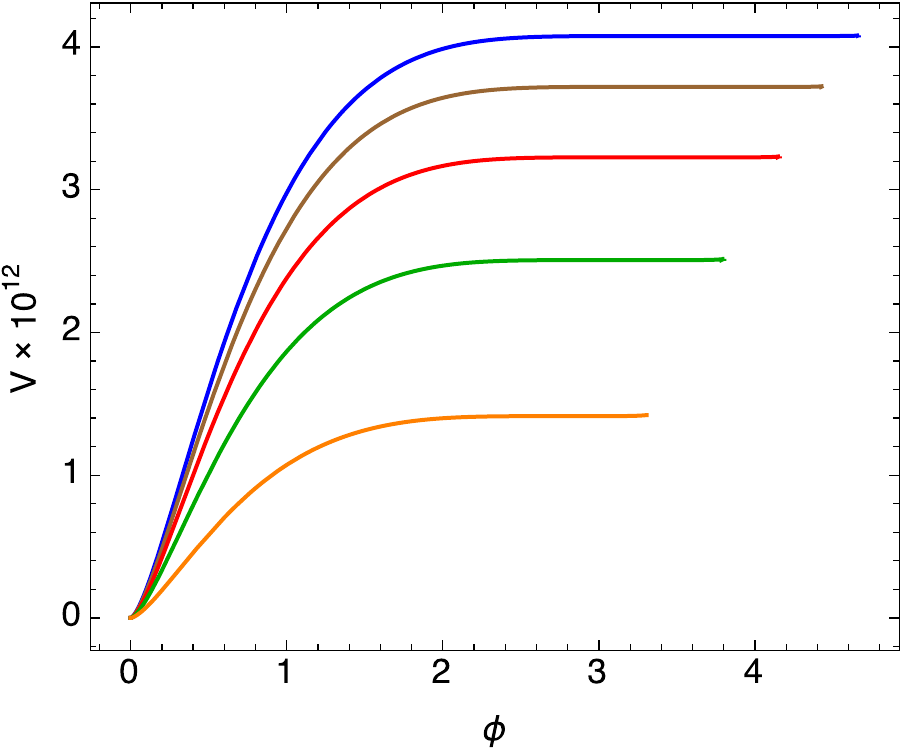}
\caption{\it Left panel: Numerical results for the model (\ref{eq:fSum}) for $N_\star = 50$ and $N_\star = 60$ (red and blue dots respectively). Right panel: Einstein frame potential for the model (\ref{eq:fSum}) for $l=6$, $l=8$, $l=10$, $l=12$ and $l=14$ (orange, green, red, brown and blue lines respectively). The saddle point lies close to the right edge of the potential, beyond which one obtains a second branch of $V$, which leads to repulsive gravity.}
\label{fig:M(l)}
\end{figure}

\begin{figure}[h]
\centering
\includegraphics[height=5.5cm]{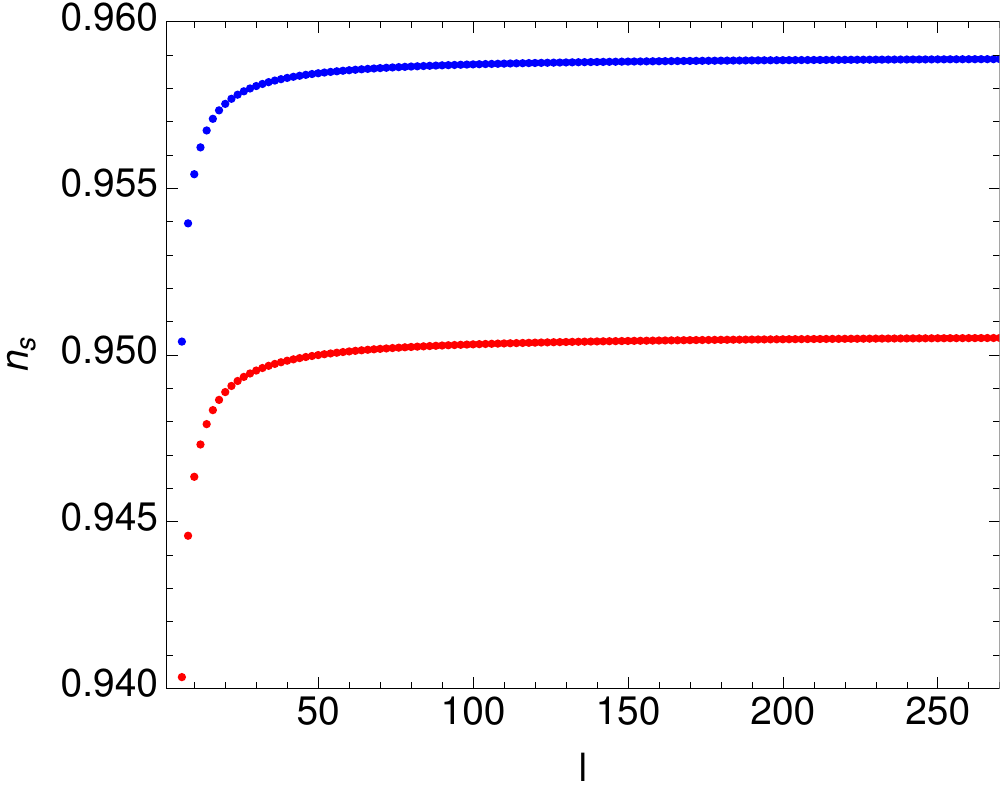}
\hspace{0.5cm}
\includegraphics[height=5.5cm]{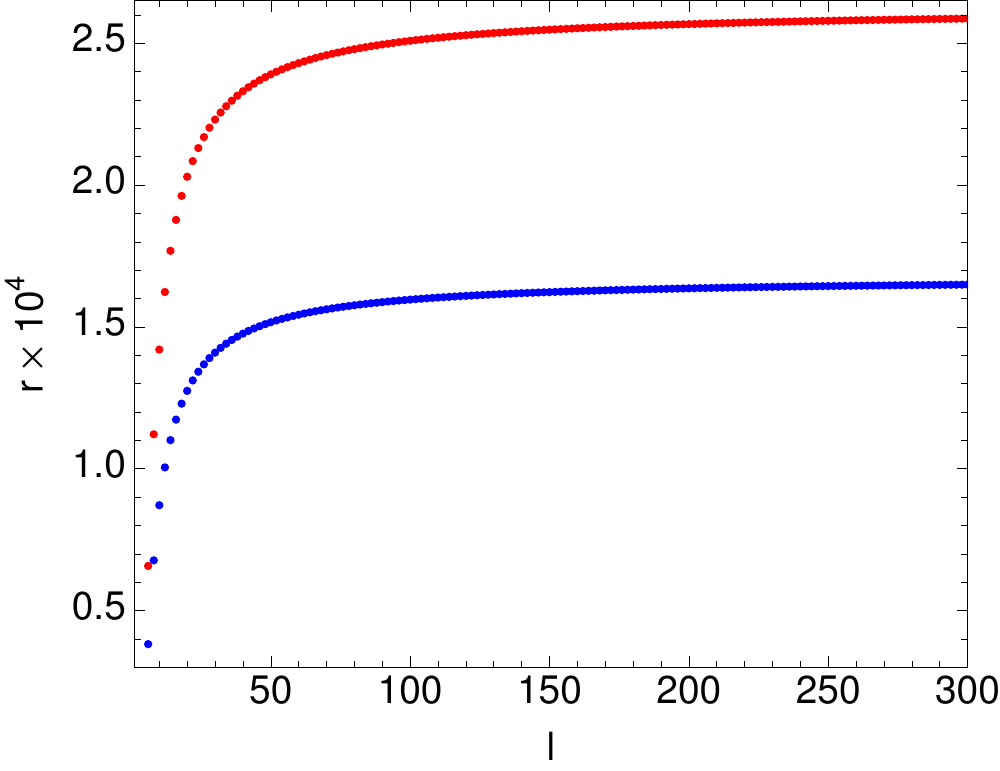}
\caption{\it Numerical results for the model (\ref{eq:fSum}) for $N_\star = 50$ and $N_\star = 60$ (red and blue dots respectively). All values of $r$ obtained in this analysis are consistent with PLANCK, but $n_s$ fits the PLANCK data only for $N_\star\simeq60$.}
\label{fig:nsr(l)}
\end{figure}

\subsection{{The} $l \to \infty$ limit} \label{sec:infterms}

\begin{figure}[h]
\centering
\includegraphics[height=5.5cm]{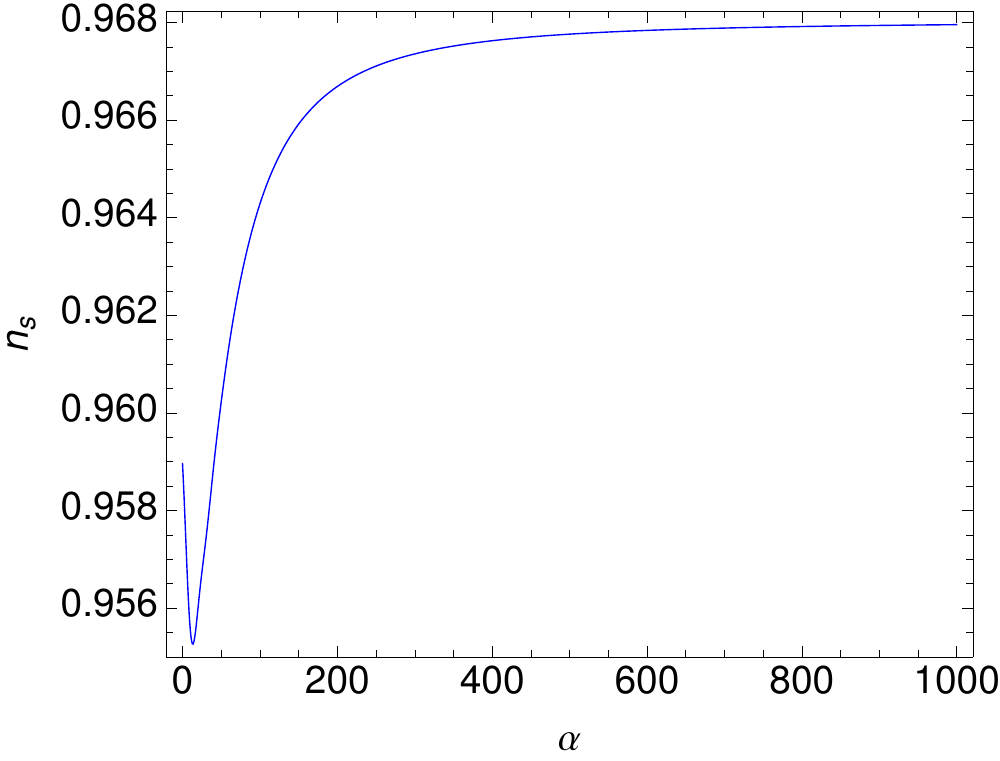}
\hspace{0.5cm}
\includegraphics[height=5.5cm]{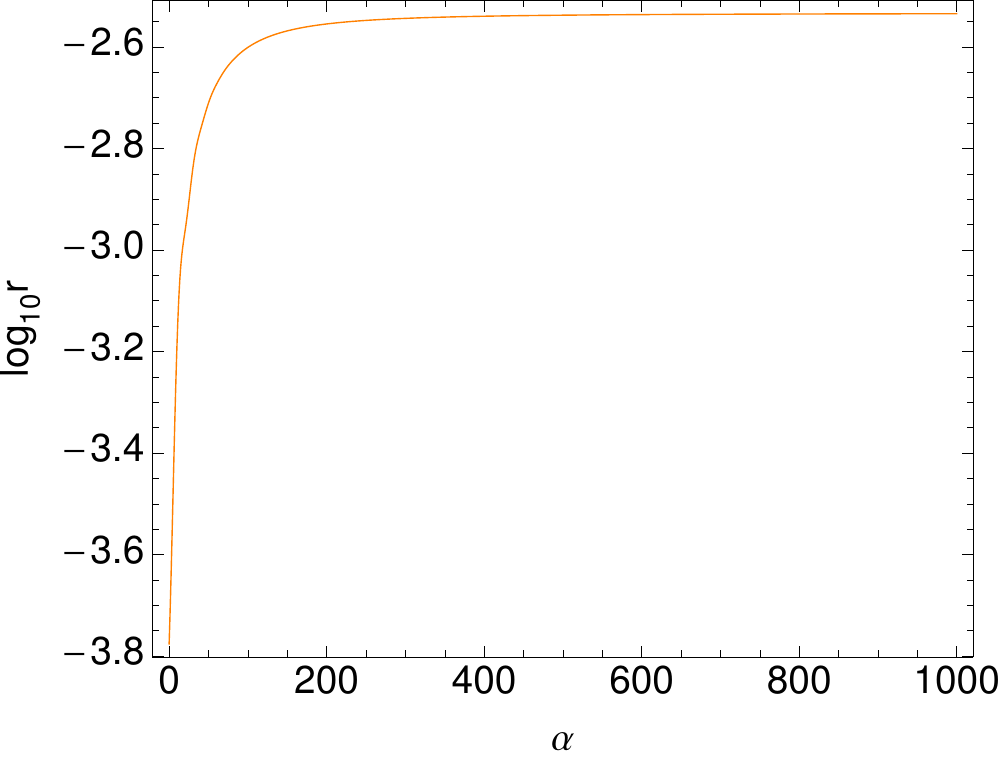}
\caption{\it Numerical results for the model (\ref{eq:fSumAll}) for $N_\star = 60$ The $n_s$ fits the PLANCK data for $0<\alpha<1.4$ and for $\alpha_2\gtrsim 34$, when the $\alpha_2$ term dominates the inflationary evolution.}
\label{fig:nsr(alpha)}
\end{figure}

\begin{figure}[h]
\centering
\includegraphics[height=5.5cm]{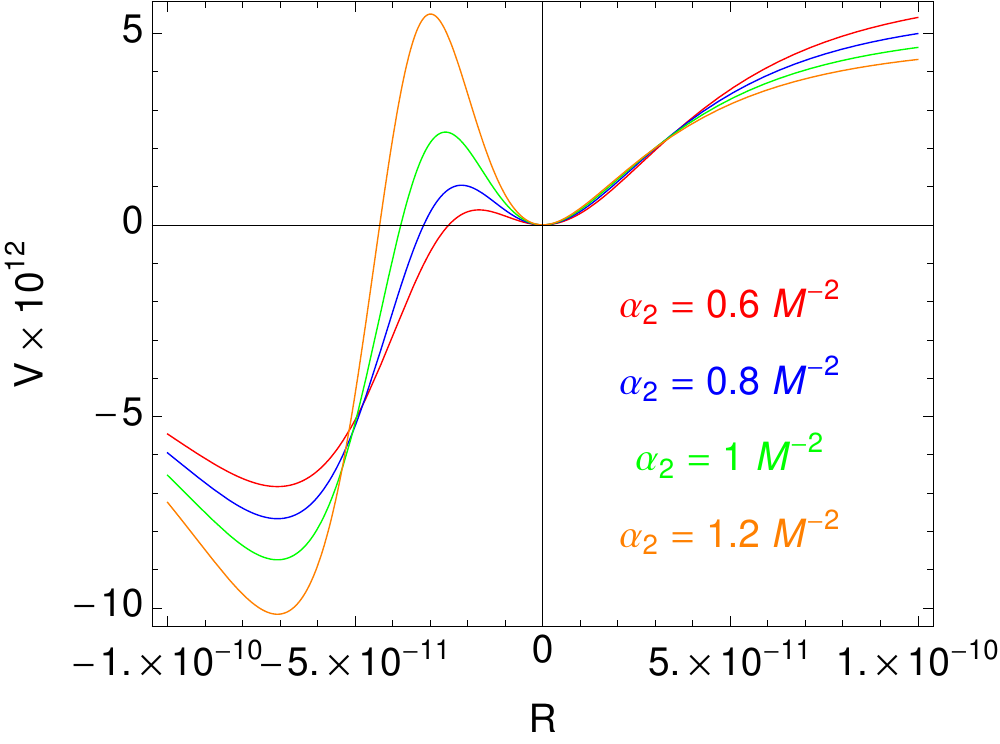}
\hspace{0.5cm}
\includegraphics[height=5.5cm]{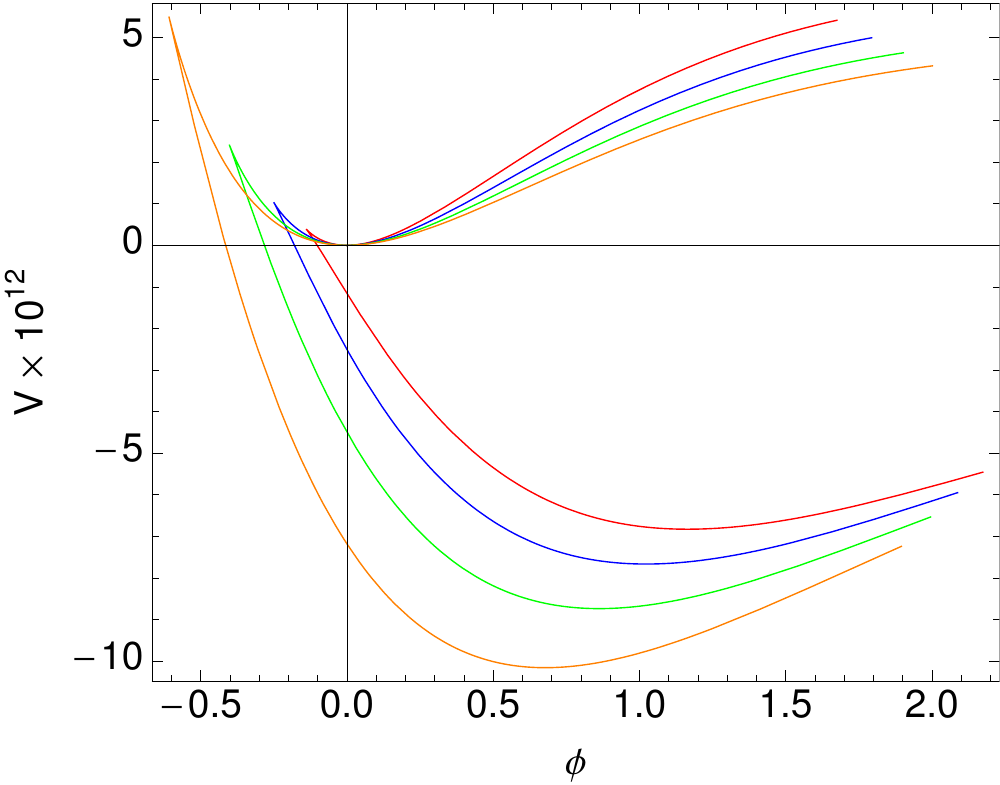}
\caption{\it Left Panel: The Einstein frame potential as a function of the Ricci scalar. The GR minimum at $R=0$ appears to be meta-stable, with a possibility of tunnelling to anti de Sitter vacuum. Right Panel: The Einstein frame potential $V$ as a function of the Einstein frame field $\phi$ for the model (\ref{eq:fSumAll}). Two branches of potential correspond to two solutions of $\varphi = F(R)$. In order to avoid overshooting the minimum at $R=0$ one requires $\alpha_2 \gtrsim 0.7$.}
\label{fig:V(alpha)}
\end{figure}

Numerical analysis shows that in order to obtain correct normalisation of primordial inhomogeneities one needs $M=M(l)$. Nevertheless for $l\to \infty$ one obtains $M \to M_o$ (where $M_o\sim 10^{-5}$ for $\alpha_2=0$), which implies ${\bf R}_s \to \infty$ for $l \to \infty$. Hence for $l \gg 1$ one cannot obtain inflation close to saddle point. For $l\to\infty$ one obtains 
\begin{equation}
f(R) = R \left(e^{-\frac{\sqrt{2} R}{M_o^2}}+\frac{\sqrt{2}+\alpha_2}{M_o^2}R\right) \, . \label{eq:fSumAll}
\end{equation}
The $\alpha_2$ may be again used to stabilise the GR vacuum at $\phi=0$. The numerical results for $N_\star = 60$ are plotted in Fig. \ref{fig:nsr(alpha)} and \ref{fig:V(alpha)}. As expected, for $\alpha \gg 1$ values of $M/\sqrt{\alpha}$, $r$ and $n_s$ obtain the limit of the Starobinksy theory. As shown in Fig. \ref{fig:V(alpha)} the potentials have two branches, which split at some $\phi = \phi_m$, where $\phi_m$ is the minimal value of $\phi$. The $\alpha_2$ term in necessary in order to stabilise the GR vacuum. For $\alpha_2 = 0$ one obtains two branches of potential which grow from $\phi = 0$. Both of them exist only for $\phi>0$ with no minimum. While increasing the value of $\alpha_2$ the splitting of branches moves towards $\phi < 0$ and the inflationary branch obtains minimum at $\phi = 0$. We investigated the stability of minimum from the perspective of classical evolution of the Einstein frame field. Namely, we considered the slow-roll initial conditions at $\phi = \phi_\star$ for different values of $\alpha_2$ and checked whether the minimum is deep enough to stop the field before it would reach $\phi_m$. {We postpone the issue of quantum tunnelling to the anti - de Sitter vacuum for future work.}

\begin{figure}[h]
\centering
\includegraphics[height=5.5cm]{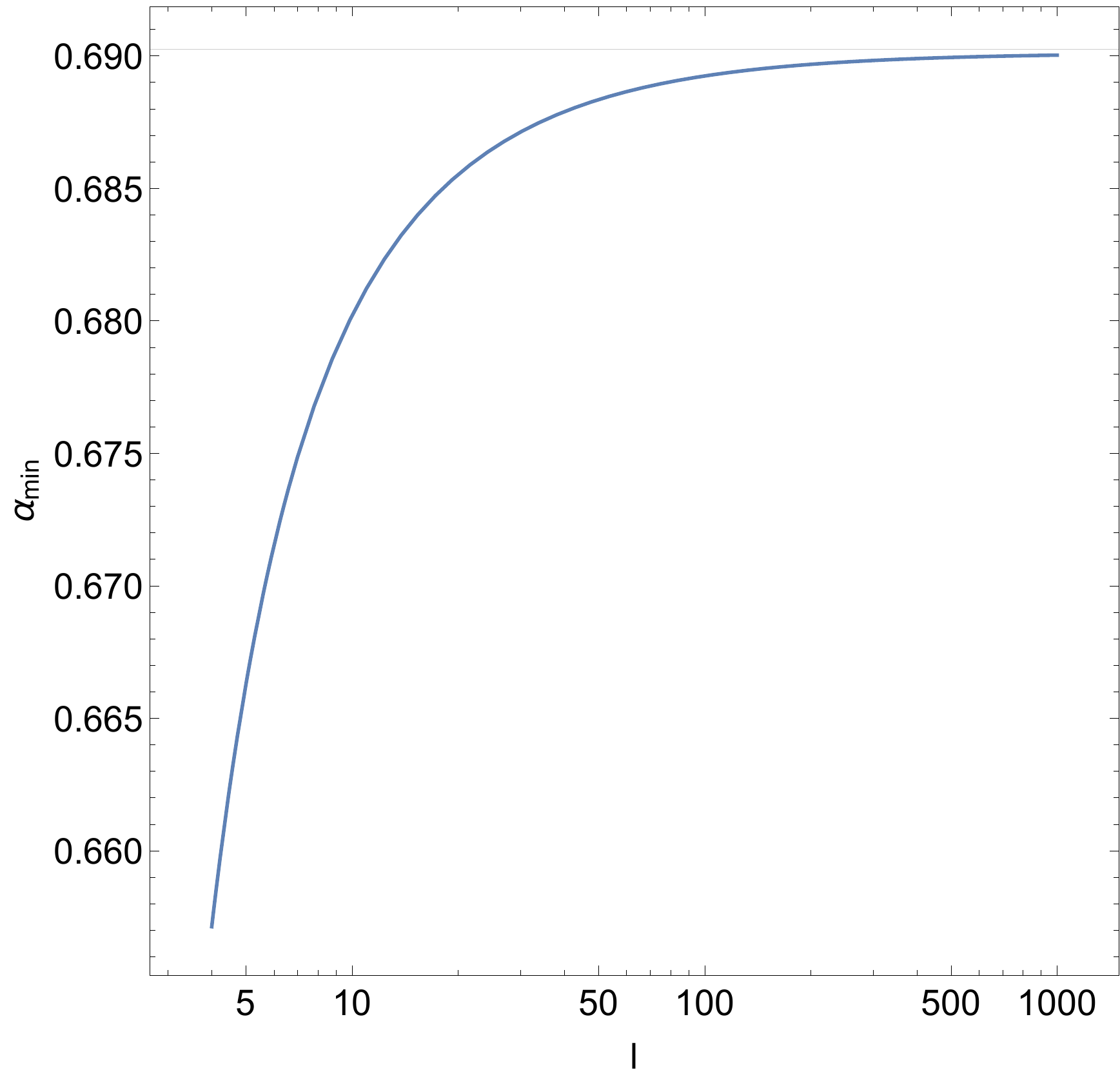}
\caption{\it The minimal value of $\alpha_2$, which allows to avoid overshooting the meta-stable GR minimum.}
\label{fig:alphallog}
\end{figure}


\section{Conclusions}\label{sec:concl}

In this paper we considered several $f(R)$ theories with saddle point in the Einstein frame potential. All models consist of GR term $R$, Starobinsky term $\alpha_2 R^2$ and higher order terms which are the source of the saddle point. In subsection \ref{sec:2terms} we investigated two additional terms proportional to $R^n$ and $R^{n+1}$. We found analytical relation between their coefficients and $R_s$, which is the value of the Ricci scalar at the saddle point. The potential {becomes unstable}
 for $R$ slightly bigger than $R_s$ - the second branch of the auxiliary field equation $\varphi = F(R)$ becomes physical, which leads to the second branch of potential and as a consequence to repulsive gravity. Significant contribution of  the $R^2$ term extend the plateau before the saddle point and pushes away the instability from the inflationary region. {For $n\geq3$ it is impossible to obtain correct $n_s$}, however for $n$ slightly bigger than $2$ one can fit the PLANCK data. 
\\*

In subsection \ref{sec:kterms} we investigated The Einstein frame potential with zero value of the first $l-2$ derivatives at the saddle point, where $l \geq 6$ is an even natural number. To obtain such a saddle point we considered $f(R)=R + \alpha_2 R^2 + \sum_{n=3}^l \alpha_n M^{2(1-n)}R^n$. We found analytical formulae for $R_s$ and for all $\alpha_n$ coefficients, as well as the explicit value of $f(R)$ after summation. {Unfortunately the result is slightly disappointing, because the saddle point moves away from the scale of freeze-out of primordial inhomogeneities with growing $l$. Thus bringing us closer to the Starbinsky case as $l$ gets bigger.} We also obtained numerical results for $n_s$, $r$ and for the suppression scale $M$ as a function of $l$. The final result strongly depends on $N_\star$, and therefore on the thermal history of the universe. One can fit the PLANCK data for $l \gtrsim 20$ and $N_\star \simeq 60$ even for $\alpha_2 = 0$. Again, for $R$ slightly bigger than $R_s$ one obtains an instability of potential, which for big $l$ is orders of magnitude away from the freeze-out scale.
\\*

In subsection \ref{sec:infterms} we {considered} the limit $l\to\infty$, which resulted in $f(R) = R (e^{-\sqrt{2} R/M_o^2}+(\sqrt{2}+\alpha_2) R/M_o^2)${, which is basically Starobinsky model plus an exponentially suppressed correction}. In such a case the saddle point (and therefore the instability for $R>R_s$) moves to infinity and inflation happens far away from the saddle point. The $\alpha_2$ term is necessary to create the meta-stable minimum of the Einstein frame potential. One can fit the PLANCK data for $0.7\lesssim\alpha_2\lesssim 1.4$ and $\alpha_2 \gtrsim 34$

{
\section*{Acknowledgements}
This work was partially supported by the Foundation for Polish Science International PhD Projects Programme co-financed by the EU European Regional Development Fund and by National Science Centre under research grants DEC-2012/04/A/ST2/00099 and DEC-2014/13/N/ST2/02712.
ML was supported by the Polish National Science Centre under doctoral scholarship number 2015/16/T/ST2/00527. MA was supported by National Science Centre grant FUGA UMO-2014/12/S/ST2/00243. 
}


\end{document}